\documentclass[]{spie}

\usepackage{amsmath,amsfonts,amssymb}
\usepackage{graphicx}
\usepackage{multirow}
\usepackage{subcaption}
\usepackage{sidecap}
\usepackage[table, x11names]{xcolor}
\usepackage[colorlinks=true, allcolors=blue]{hyperref}

\title{Liger for Next Generation Keck AO: Filter Wheel and Pupil Design}

\author[a,b]{Maren Cosens}
\author[a,b]{Shelley A. Wright}
\author[c,d]{Pauline Arriaga}
\author[b]{Aaron Brown}
\author[c]{Michael Fitzgerald}
\author[e]{Tucker Jones}
\author[f]{Marc Kassis}
\author[c]{Evan Kress}
\author[g]{Renate Kupke}
\author[c]{James E. Larkin}
\author[f]{Jim Lyke}
\author[c]{Eric Wang}
\author[a,b]{James Wiley}
\author[f]{Sherry Yeh}
\affil[a]{Department of Physics, University of California San Diego, USA}
\affil[b]{Center for Astrophysics and Space Sciences, University of California San Diego, USA}
\affil[c]{Department of Physics \& Astronomy, University of California Los Angeles, USA}
\affil[d]{Areté Associates, Northridge, CA}
\affil[e]{Department of Physics, University of California Davis, USA}
\affil[f]{W.M. Keck Observatory, Waimea, HI}
\affil[g]{Department of Astronomy \& Astrophysics, University of California Santa Cruz, USA}

\authorinfo{Further author information: (Send correspondence to M.C.)\\M.C.: E-mail: mcosens@ucsd.edu}
 
\begin{document} 
\maketitle

\begin{abstract}
Liger is a next-generation near-infrared imager and integral field spectrograph (IFS) for the W.M. Keck Observatory designed to take advantage of the Keck All-Sky Precision Adaptive Optics (KAPA) upgrade. Liger will operate at spectral resolving powers between R$\sim$4,000 - 10,000 over a wavelength range of 0.8-2.4$\mu$m. Liger takes advantage of a sequential imager and spectrograph design that allows for simultaneous observations between the two channels using the same filter wheel and cold pupil stop. We present the design for the filter wheels and pupil mask and their location and tolerances in the optical design. The filter mechanism is a multi-wheel design drawing from the heritage of the current Keck/OSIRIS imager single wheel design. The Liger multi-wheel configuration is designed to allow future upgrades to the number and range of filters throughout the life of the instrument. The pupil mechanism is designed to be similarly upgradeable with the option to add multiple pupil mask options. A smaller wheel mechanism allows the user to select the desired pupil mask with open slots being designed in for future upgrade capabilities. An ideal pupil would match the shape of the image formed of the primary and would track its rotation. For different pupil shapes without tracking we model the additional exposure time needed to achieve the same signal to noise of an ideal pupil and determine that a set of fixed masks of different shapes provides a mechanically simpler system with little compromise in performance.
\end{abstract}


\section{INTRODUCTION}\label{sec:intro}
Liger is a next generation adaptive optics-fed integral field spectrograph (IFS) and imager being designed to utilize the upcoming Keck All-Sky Precision Adaptive Optics (KAPA) upgrade. Liger will provide improvements over existing instruments, operating at spectral resolving powers up to R$\sim$4,000-10,000 over a wavelength range of 0.8-2.4$\mu$m. The Liger imager will provide a 10mas per pixel plate scale with a field of view of 20.4x20.4 arcseconds. The design of the Liger imager draws on the heritage of the existing OSIRIS imager\cite{Larkin2006, Arriaga2018} at Keck while the spectrograph is adapted from the extensive design work done for IRIS\cite{Larkin2016}, the first light instrument for the Thirty Meter Telescope (TMT). The Liger IFS will be a duplicate of IRIS and become a pathfinder instrument for it's development, while the imager is being custom designed for Liger. A more complete overview of the Liger instrument is provided in Wright et al. (\textit{this conference})\cite{Wright2020}.

Here we present the design of the Liger filter wheel and pupil stop. The imaging camera serves as the reimaging optical system for the spectrographs, so these elements (as well as the rest of the imager optics) will be simultaneously used by the spectrograph. This allows for improved AO correction in the imager as well as improvements in masking the thermal background for the IFS by making use of the larger pupil located in the imager.

\section{FILTER WHEEL}\label{sec:fwheel}
\begin{figure}
\begin{subfigure}{0.36\textwidth}
    \centering
    \includegraphics[width=\textwidth]{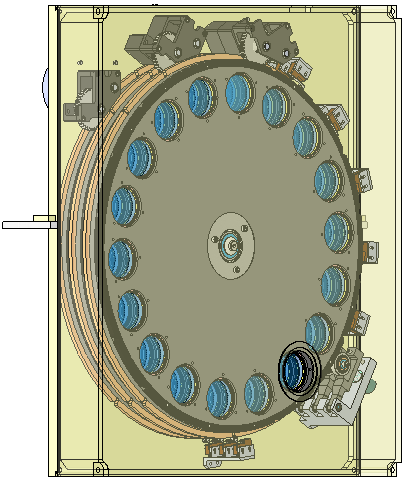}
    \caption{}
\end{subfigure}
\begin{subfigure}{0.36\textwidth}
    \centering
    \includegraphics[width=\textwidth]{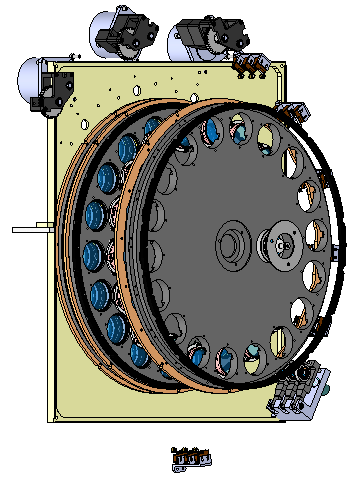}
    \caption{}
\end{subfigure}
\begin{subfigure}{0.26\textwidth}
    \centering
    \includegraphics[width=\textwidth]{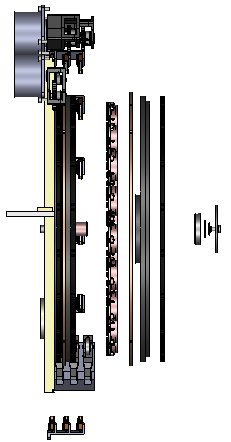}
    \caption{}
\end{subfigure}
\caption{Model Views of the full Liger filter wheel assembly. (a): Full assembly with transparent housing to show internal components. The motors for each wheel are mounted at the top of the housing on the opposite side with individual spur gear assemblies (see Figure \ref{fig:gears}) interfacing with the wheels. A stack of three detents are located in the bottom right (see Figure \ref{fig:detent}). These detents will provide enough force to hold the filter wheel in a stable position during operation of the imager. Along the right side and bottom of the filter wheel are six stacks of switches which will indicate the position of each wheel. (b,c): Partially exploded views of the filter wheel assembly showing the separation of one of the main wheels (gray), its switch activator (brown) and detent rings (black), filter sub-assemblies (burnt-red and gray, see Figure \ref{fig:filter_ass}), and mounting components. Switches, detents, motors, and gearing are exploded radially outward. Some mounting screws are excluded for improved visibility of components. \label{fig:full_model}}
\end{figure}

The design of the filter wheel draws from the heritage of the OSIRIS imager upgrade\cite{Arriaga2018}, but is expanded to provide a significant increase in the number of available filter slots. In order to provide a sufficient number of filter slots and fit the Liger dewar volume, we make use of three stacked filter wheels. Each wheel has it's own dedicated motor, switches, and detent, but follow the same design. A model of this three wheel design is shown in Figure \ref{fig:full_model}.

Each wheel consists of 18 filter slots for a total of 51 available slots (1 slot per wheel will need to be a clear aperture). The front wheel will be located 50mm behind the pupil plane at which point the beam will be 25.8mm in diameter. The exit of the last filter wheel is 56mm behind this point. The beam diameter by this point will be $\sim$30mm. We will be using 1.5in (38.1mm) filters mounted with a retainer of diameter 35.7mm. This leaves 2.85mm radially for tolerance in positioning (with greater tolerance at the first two wheels) before vignetting will occur. The stepper motors being used to position the wheels have a step angle of 1.8$^\circ$ and a step accuracy of 3\%. With the gear ratios discussed in Section \ref{sec:motor_gears} and filters located at a radius of 145mm, this results in a positioning accuracy of 0.2mm radially from the center of the optical beam. There then remains 2.65mm in tolerance for the manufacturing and positioning of the wheels.

\begin{figure}[h]
\begin{subfigure}{0.33\textwidth}
    \centering
    \includegraphics[scale=0.33]{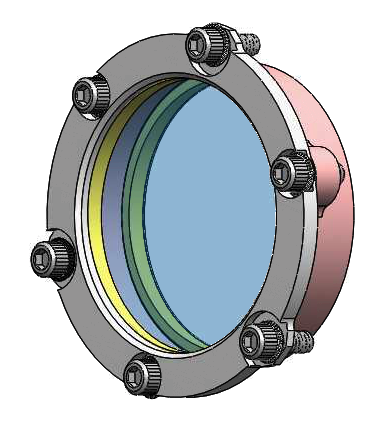} 
    \caption{}
    \label{fig:filter_ass}
\end{subfigure}
\begin{subfigure}{0.33\textwidth}
    \centering
    \includegraphics[scale=0.25]{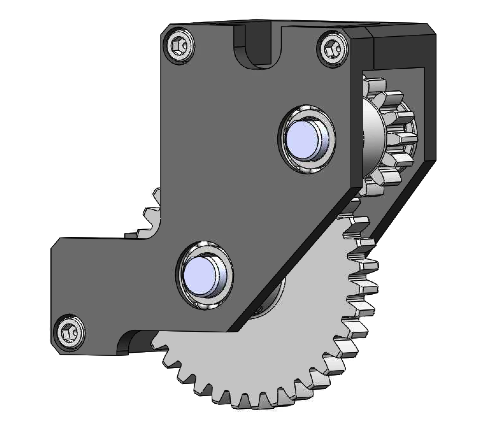}
    \caption{}
    \label{fig:gears}
\end{subfigure}
\begin{subfigure}{0.33\textwidth}
    \centering
    \includegraphics[scale=0.25]{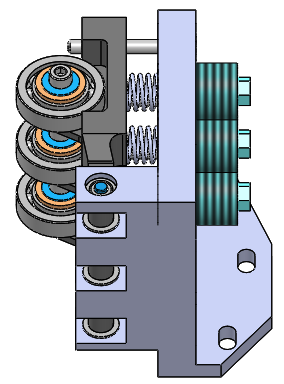}
    \caption{}
    \label{fig:detent}
\end{subfigure}
\begin{subfigure}{0.33\textwidth}
    \centering
    \includegraphics[scale=0.25]{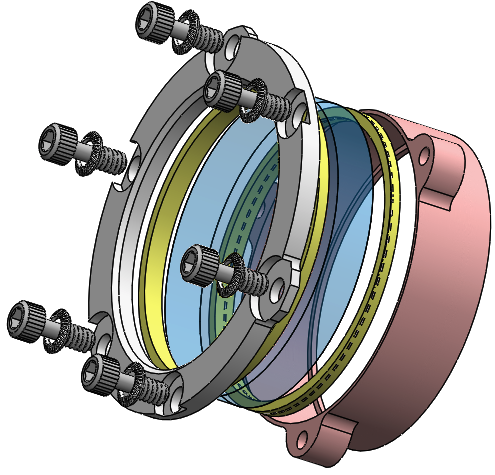} 
    \caption{Filter Assembly}
    \label{fig:filter_ass2}
\end{subfigure}
\begin{subfigure}{0.33\textwidth}
    \centering
    \includegraphics[scale=0.3]{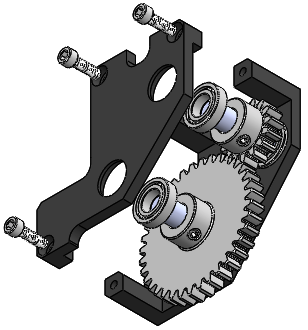}
    \caption{Spur gear assembly}
    \label{fig:gears2}
\end{subfigure}
\begin{subfigure}{0.33\textwidth}
    \centering
    \includegraphics[scale=0.3]{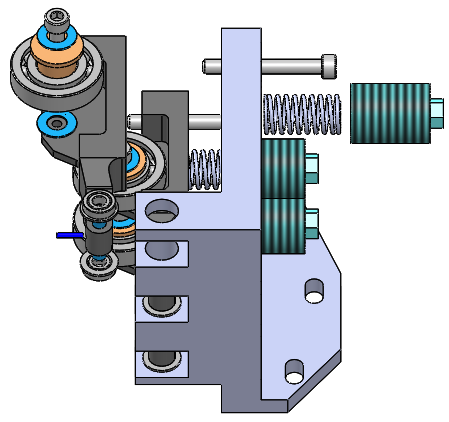}
    \caption{Detent assembly}
    \label{fig:detent2}
\end{subfigure}
\caption{Close up views of key components of the Liger filter wheel. (a,d): Mounting assembly for individual filters; this is the same design currently used in the OSIRIS imager expanded for larger 1.5 in filters. Spacers (yellow) are placed on either side of the 1.5\,in diameter filter (transparent blue) inside the housing. These are held in by the retainer ring (silver) which also secures the assembly in the filter wheel. (b,e): Spur gear assembly which will be the interface between the stepper motors and the filter wheels. The pinion will be directly coupled to the stepper motor shaft with the larger gear mated to the gear teeth on the filter wheel. Ball bearings are pressed into the housing (dark grey) on each side of the assembly to support the shafts. Since the three filter wheels are stacked, aluminum spacers will be used to set the height of two of these assemblies. (c,f): Detent assembly with three independent arms to maintain accurate positioning of the filter wheels. The spring connected to each detent arm will be compressed by at least $0.08\,\rm in$ via the threaded spring housing (dark green) on the back of the detent in order to provide the proper amount of force to prevent the wheel from slipping. The roller bearings are where contact will be made with the filter wheels so that they are able to rotate when the stepper motors are active.}
\end{figure}

The filters will be held and mounted to the wheels using the same method as the OSIRIS imager, with the housing simply expanded for the larger diameter filters used in Liger. Each filter will be placed in a cylindrical housing sandwiched between 3.5mm spacers. This complete filter assembly is shown in Figure \ref{fig:filter_ass} and will be mounted to the filter wheels as a complete unit.

The position of each wheel is determined by a set of six binary switches, with each filter position corresponding to a unique combination of open and closed switches. A corresponding ``switch activator ring" is attached to each wheel with slots for the needed ``open" positions.

\subsection{Motor and Gearing}\label{sec:motor_gears}
If we assume a filter change should be completed in 10\, seconds, then we would require a rotation speed of $\sim$6 rpm at the filter wheel. Generally, the rotation speed from the motor is dependent on the current provided and torque required, but this is a low torque application that should not put any significant strain on the motor. As a starting point to design a gear train we assume a nominal output speed of 450 rpm. This requires a significant speed reduction and gear ratio of 75:1, requiring more than a single interface. We therefore design a simple assembly making use of a 15 tooth pinion and a 36 tooth idler gear. The idler gear then interfaces with the 334 tooth filter wheel. We check for interference making use of the following equation\cite{shigleys} for the minimum number of teeth needed on the pinion:
\begin{equation} \label{eqn:interference}
    N_p = \frac{2k}{(1+2m)sin^2\phi}\left (m+\sqrt{m^2+(1+2m)sin^2\phi} \right )
\end{equation}
where $k=1$ for full-depth teeth, $\phi$ is the pressure angle (20$^\circ$), and $m=N_G/N_p$, the ratio of teeth on the gear and pinion. The interface between the pinion and idler gear gives a minimum number of teeth on the pinion of 14, so the 15 tooth pinion will not lead to interference. For the second interface between the idler gear and filter wheel the minimum number of teeth on the idler gear is 16, so the 36 tooth idler gear does not lead to interference here either. If we were to remove the idler gear, we would have $m=N_G/N_p=334/15$ resulting in a minimum of 17 teeth on the pinion to avoid interference. Therefore, the idler gear is required to maximize the speed reduction and avoid interference. It should be noted that this only achieves a gear reduction of $\sim$22:1, but further reduction would require a more complex mechanism to avoid interference. With this ratio the motor would need to be driven at 140 rpm in order to achieve a wheel rotation speed of 6 rpm.

The assembly of these spur gears is shown in Figure \ref{fig:gears} and included in the full filter wheel assembly of Figure \ref{fig:full_model}. For the two wheels further from the mounting plate, simple spacers and shaft couplings are used to align and mate the gears with the motor and filter wheels.

\subsection{Detent}\label{sec:detent}
\begin{figure}[h]
    \centering
    \includegraphics[scale=0.4]{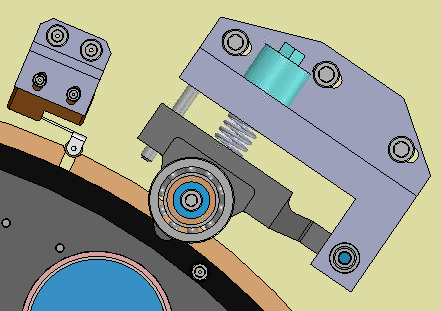}
    \caption{Filter wheel model showing the interface between the wheel and the detents ans well as one set of position switches. The detent is planar with the black catch ring which will be attached to the main wheel. This ring has notches cut into the perimeter which correspond with each filter position. The bearing on the detent arm will rest in this notch with the spring providing enough force to prevent the wheel from slipping when holding a set position. The switches are planar with the brown ring located behind the main wheel in this image. The notches cut into this ring provide the switch ``open" position. }
    \label{fig:detent_wheel}
\end{figure}

In order to maintain stable positioning of the filter wheel after moving to the correct slot, as well as repeatability over time, we make use of a detent and corresponding catch ring for each filter wheel. The mechanism consists of a lever arm pinned at one end with a spring attached near the other end pushing the arm into the catch ring. A ball bearing is attached to the lever arm providing the physical interface to the catch ring in order to allow the filter wheel to rotate when the stepper motor is activated. The detent assembly consists of three lever arms and springs which independently interface with the filter wheels. In order to fit a stack of three detents without interference with the other wheels, bearings 1/4in larger in diameter were required than in the OSIRIS counterpart. This increased the space radially between the point of contact with the wheel and the rest of the detent assembly, allowing the switch activator rings of each filter wheel to pass between the separated bearings while keeping the design of the detent arms consistent with the OSIRIS imager upgrade. A model of the stacked detent is shown in Figure \ref{fig:detent}, and the interface between this mechanism and the filter wheels is shown in Figure \ref{fig:detent_wheel}.

The spring selected for the detent mechanism must provide enough force to the filter wheel to prevent slipping but not forward driving from the motor. In order to determine the range of potential springs we first calculate the amount of torque needed to prevent the filter wheel from slipping due to mass asymmetries. Based on a solid model of the rotating components of the filter wheel the center of mass is located at a distance of $0.27\,\rm in$ from the center. This produces a maximum torque of $29\,\rm oz*in$ when it is at 90$^{\circ}$.

A standard spring should not reasonably prevent the stepper motor from moving the filter wheel, but we calculate this upper limit as a sanity check. At the motor speed of 140\,rpm determined in Section \ref{sec:motor_gears}, the output torque should be $\rm\sim31\,oz*in$. The second spur gear in the assembly described above is an idler gear so the torque transmitted to the filter wheel is simply dependent on the ratio of the filter wheel and pinion radii: $\tau_{wheel} = \frac{r_{wheel}}{r_p} \tau_{motor}$. This results in $\tau_{wheel} \sim 690\,\rm oz*in$; setting the upper limit of the torque that could be applied by the detent while still allowing rotation of the filter wheel by the motor.

Based on the location of the detent and a spring compression of $0.1\,\rm in$, this results in a required spring constant, k, between $5.5\, - \,134.7\, \rm lb_f\,in^{-1}$. We therefore choose a spring with $\rm k \, = \, 6.92\, lb_f \, in^{-1}$, requiring $\sim0.8\, \rm in$ of compression. As expected, this will not interfere with movement of the filter wheels when the stepper motors are active.

\section{PUPIL WHEEL}\label{sec:pupil}
\begin{SCfigure}[\sidecaptionrelwidth][h]
\centering
\begin{subfigure}{0.245\textwidth}
    \centering
    \includegraphics[width=.8\textwidth]{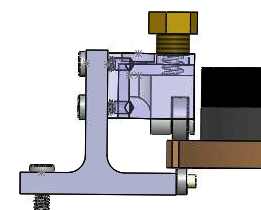}
    \caption{}
\end{subfigure}
\begin{subfigure}{0.245\textwidth}
    \centering
    \includegraphics[scale=0.25]{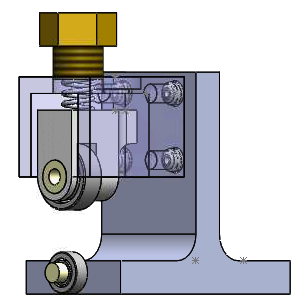}
    \caption{}
\end{subfigure}
    \caption{Spring loaded bearing mechanism to maintain planarity of the Liger pupil wheel. The spring at the top will be compressed to provide downward force through the upper bearing to the switch activator ring (brown) which extends radially outward from the pupil wheel (gray). Three of these mechanisms will be located symmetrically around the pupil wheel.}
    \label{fig:pupil_bearing}
\end{SCfigure}

Unlike our predecessor, the OSIRIS imaging camera, the Liger pupil masks will be held in a dedicated mechanism at the pupil plane rather than being included with the filter wheel.
The pupil wheel is in the early stages of the design, but will nominally consist of a wheel with slots for seven unique pupil masks. The overall design of the pupil wheel will mirror that of the filter wheel; making use of the same motor and switches along with a similar detent mechanism. However, the positioning of the pupil masks requires higher precision than the filters and therefore some additional components will be needed. In addition to a detent arm keeping the wheel from slipping, the wheel will be pinched between spring loaded bearing mechanisms at three locations in order to keep it planar (see Figure \ref{fig:pupil_bearing}). At each of these locations there will be a ball bearing at a fixed height on the back side of the wheel, with a second ball bearing attached to a compression spring on the front side. This spring will be adjusted in order to provide enough force to maintain planarity of the pupil wheel.

A set of standard masks are currently planned to be used in conjunction with the pupil wheel following the inscribed circle, hexagonal, and matched hex shapes, optimized to the short and long ends of the Liger wavelength coverage, as well as specialty masks such as a vector Apodizing Phase Plate (vector-APP) coronagraph. This unique type of coronagraph is designed for use at the pupil plane to provide an increase in contrast for directly imaging exoplanets by creating a dark hole in the point-spread-function (PSF) of the host star\cite{Snik2012}.

The use of a dedicated mechanism to house the pupil masks allows the possibility of adding rotating masks rather than fixing their orientation. This would add significant complexity to the design, so we therefore performed a trade study to determine the signal-to-noise gains of this functionality.

\subsection{Pupil Simulations}\label{sec:pupil_sims}
We simulate the background noise difference between using a mechanically fixed pupil mask and a rotating pupil to match the field rotation on sky. We use the Keck pupil plane simulations from Arriaga et al. (2016)\cite{Arriaga16} which give a model of the throughput and background emission as a function of position as the basis for this study. These models were developed using images from the pupil-viewing mode of NIRC2 with the Kp ($\rm \lambda \sim 2.124\mu m$), PAH ($\rm \lambda \sim 3.290\mu m$), and Br-$\alpha$ ($\rm \lambda \sim 4.052\mu m$) filters; we limit this analysis to the Kp filter as PAH and Br-$\alpha$ fall outside the wavelength coverage of Liger.

The pixel scale was updated to the 10mas scale of the Liger Imager. Masks were generated in the large-hex, inscribed circle, and matched hex shapes to calculate the Signal-to-Noise Ratio (SNR) with different mask configurations. The dimensions of the masks were also taken from the optimization of Arriaga et al. (2016)\cite{Arriaga16}. The SNR was calculated with a mechanically fixed mask (giving relative rotation between the image pupil and the mask) as well as a rotating pupil mask which would counter this relative motion keeping the orientation of the image pupil and the mask fixed. The ``signal" in these simulations is simply a value of 1$\times$(exposure time) wherever the mask is not present. Therefore the relative SNR between the rotating and fixed pupils is the important parameter. In order to quantify this we evaluate the additional integration time needed with a mechanically fixed mask to reach the same SNR as a mask that rotates. This additional time is reported in Table \ref{tbl:backlim_snr} for a reference exposure time of 600s. In order for this to be an accurate representation of the difference between the two scenarios we take into account the affects of pupil mask misalignment and pupil nutation which we discuss in more detail below.

\begin{table}
\centering
\caption{Additional exposure time needed with a fixed mask to reach the SNR of a 600s integration with an ideal rotating pupil mask. \label{tbl:backlim_snr}}
\begin{tabular}{|c|c|c|} 
\hline
Mask & Elevation ($^\circ$) & Additional Exposure Time (s) \\
\hline
\multirow{4}{4em}{inscribed circle} & \cellcolor{Snow4} 20 & \cellcolor{Snow4} 15 \\
 & \cellcolor{Snow3} 30 & \cellcolor{Snow3} 15 \\
 & \cellcolor{Snow2} 50 & \cellcolor{Snow2} 15 \\
 & \cellcolor{Snow1} 80 & \cellcolor{Snow1} 20 \\
\hline
\multirow{4}{4em}{hex} & \cellcolor{Snow4} 20 & \cellcolor{Snow4} 15\\
 & \cellcolor{Snow3} 30 & \cellcolor{Snow3} 15 \\
 & \cellcolor{Snow2} 50 & \cellcolor{Snow2} 15 \\
 & \cellcolor{Snow1} 80 & \cellcolor{Snow1} 35 \\
\hline
\multirow{4}{4em}{matched} & \cellcolor{Snow4} 20 & \cellcolor{Snow4} 25\\
 & \cellcolor{Snow3} 30 & \cellcolor{Snow3} 15 \\
 & \cellcolor{Snow2} 50 & \cellcolor{Snow2} 20 \\
 & \cellcolor{Snow1} 80 & \cellcolor{Snow1} 50 \\
\hline
\end{tabular}
\end{table}

\begin{figure}[h]
    \centering
    \includegraphics[width=0.6\textwidth]{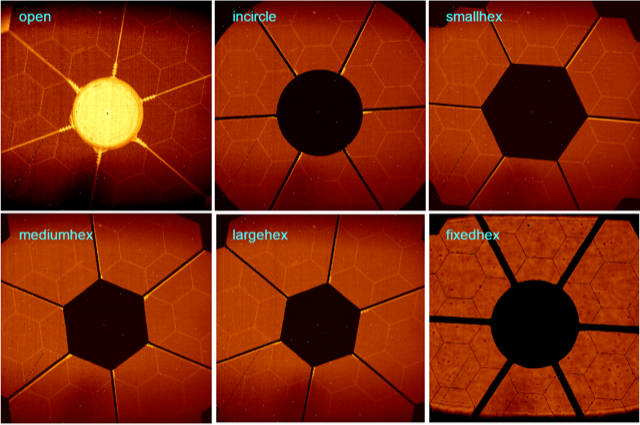}
    \caption{Images taken in March 2019 with the NIRC2 pupil imaging camera in the Kp filter after the most recent upgrade to the instrument showing the relative alignment between the pupil stop and the telescope pupil. All images were taken with the dome closed except for the bottom right. We match the offset seen here to define the alignment accuracy of the pupil masks in our simulations.}
    \label{fig:nirc2_pupil}
\end{figure}

In images taken with the NIRC2 pupil viewing camera (Figure \ref{fig:nirc2_pupil}), it is clear that there is a slight  misalignment between the mask and the pupil image. This misalignment is expected to be present in any system due to the very fine precision needed to perfectly align the mask and to then keep it aligned over time while moving different masks in and out of the beam (particularly if the mask is to rotate). We add a conservative relative offset of 4.5cm at the plane of the primary mirror between the mask and the pupil image in our simulations to match that seen in Figure \ref{fig:nirc2_pupil}. This is equivalent to a 0.1mm misalignment at the pupil plane. The real system may have a larger offset than this over time after repeated cycles of changing pupil masks.

\begin{figure}[h]
\centering
\begin{subfigure}{0.245\textwidth}
    \centering
    \includegraphics[scale=0.25]{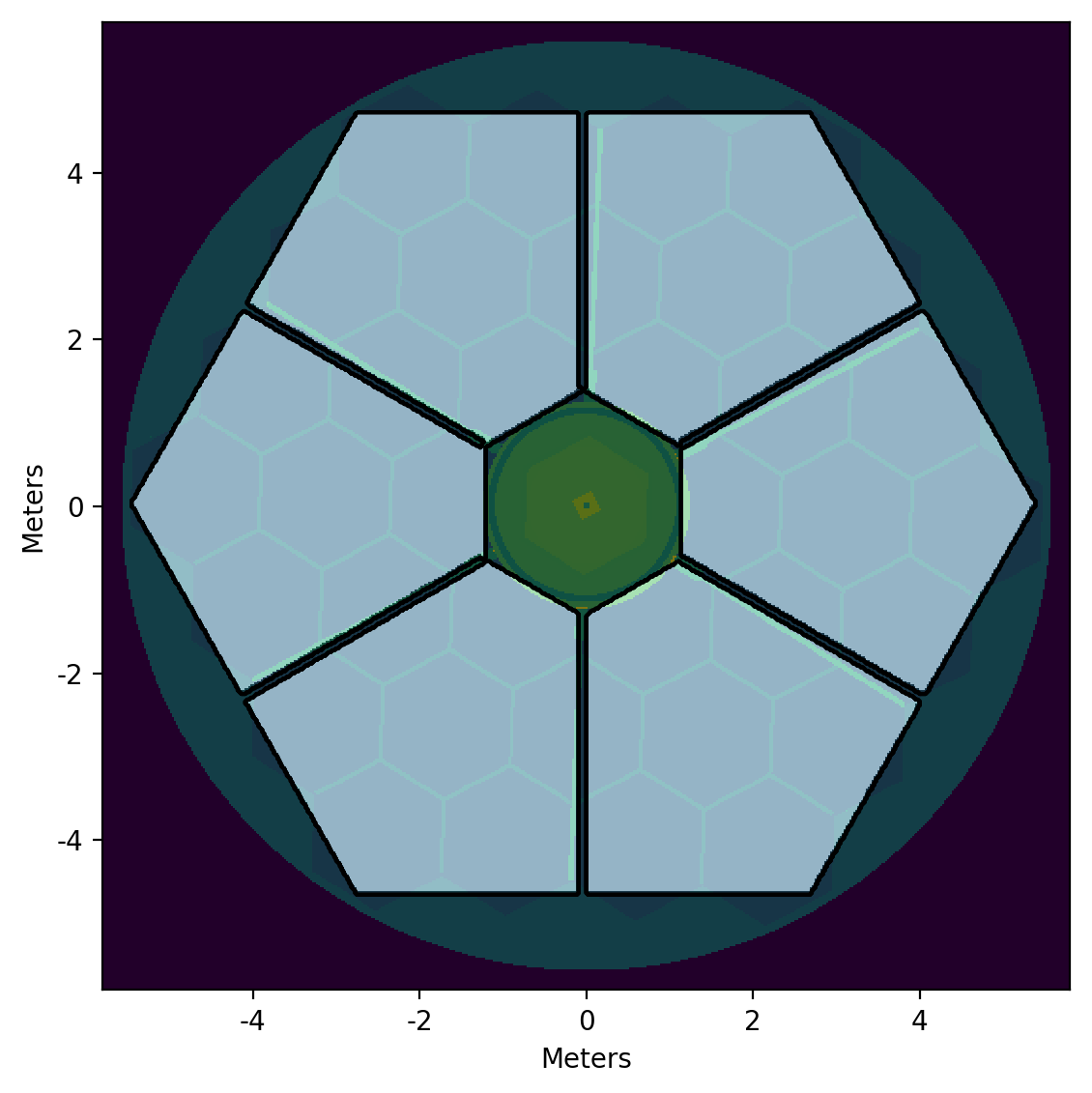}
    \caption{20$^{\circ}$}
\end{subfigure}
\begin{subfigure}{0.245\textwidth}
    \centering
    \includegraphics[scale=0.25]{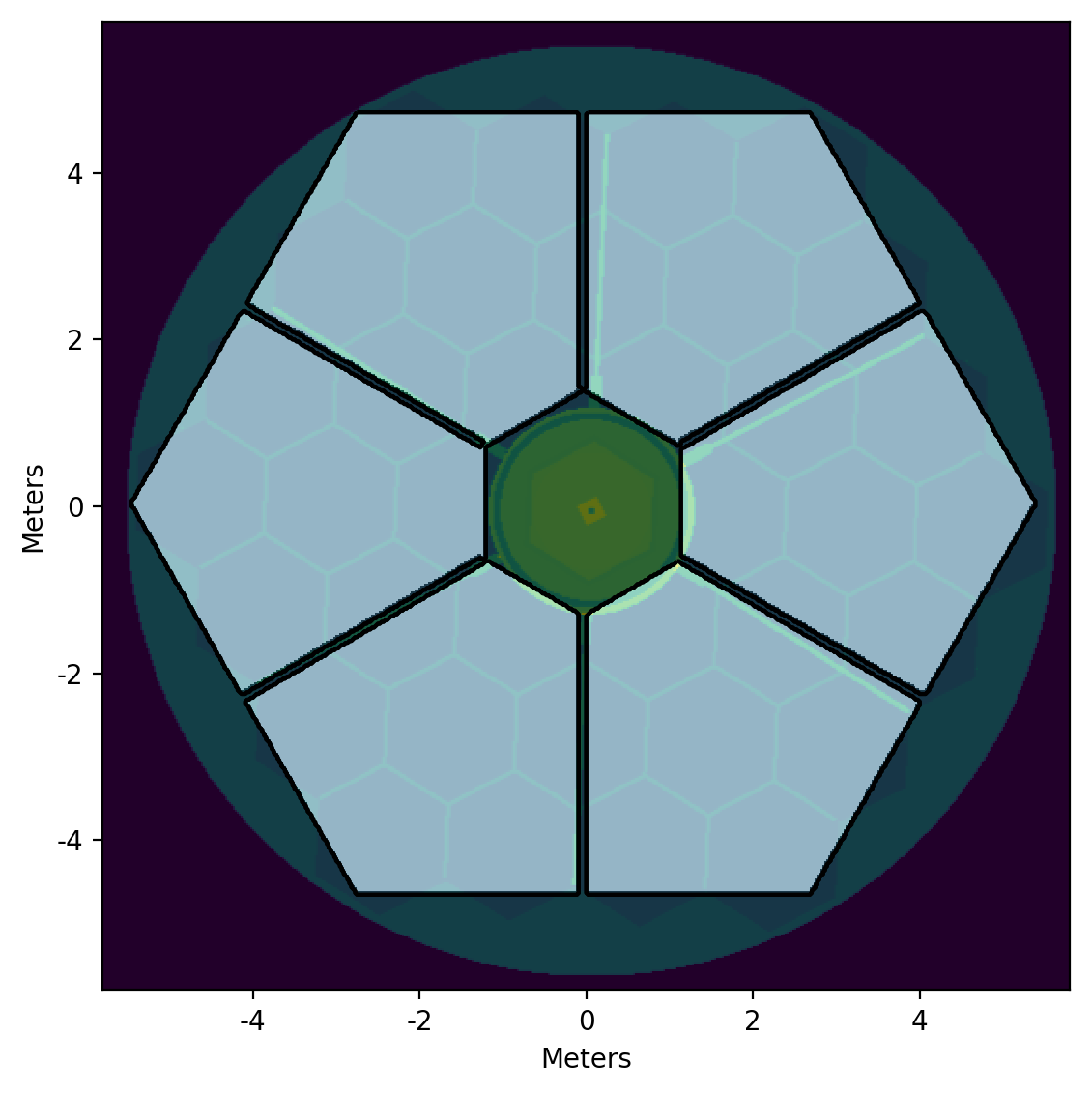}
    \caption{30$^{\circ}$}
\end{subfigure}
\begin{subfigure}{0.245\textwidth}
    \centering
    \includegraphics[scale=0.25]{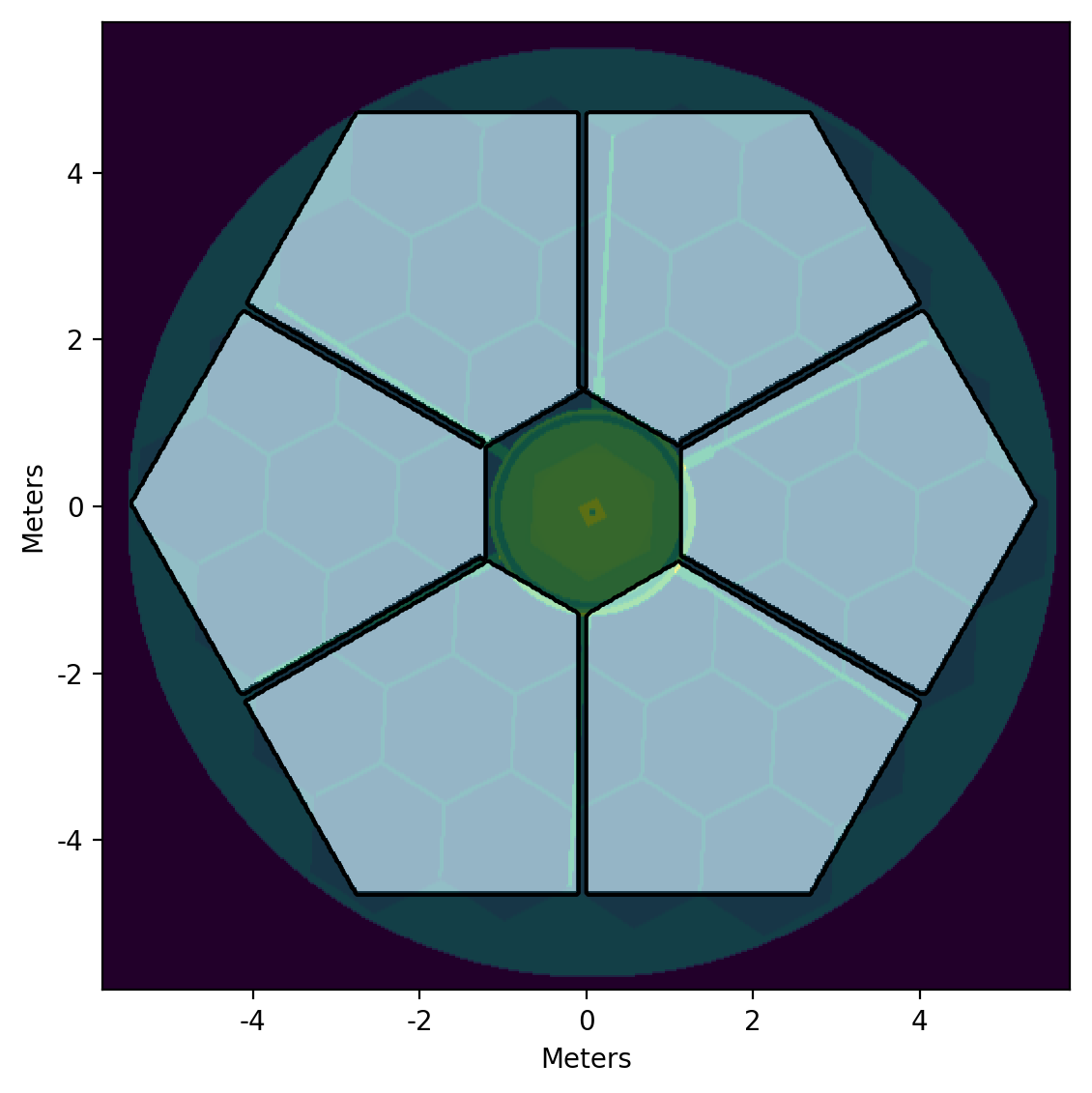}
    \caption{50$^{\circ}$}
\end{subfigure}
\begin{subfigure}{0.245\textwidth}
    \centering
    \includegraphics[scale=0.25]{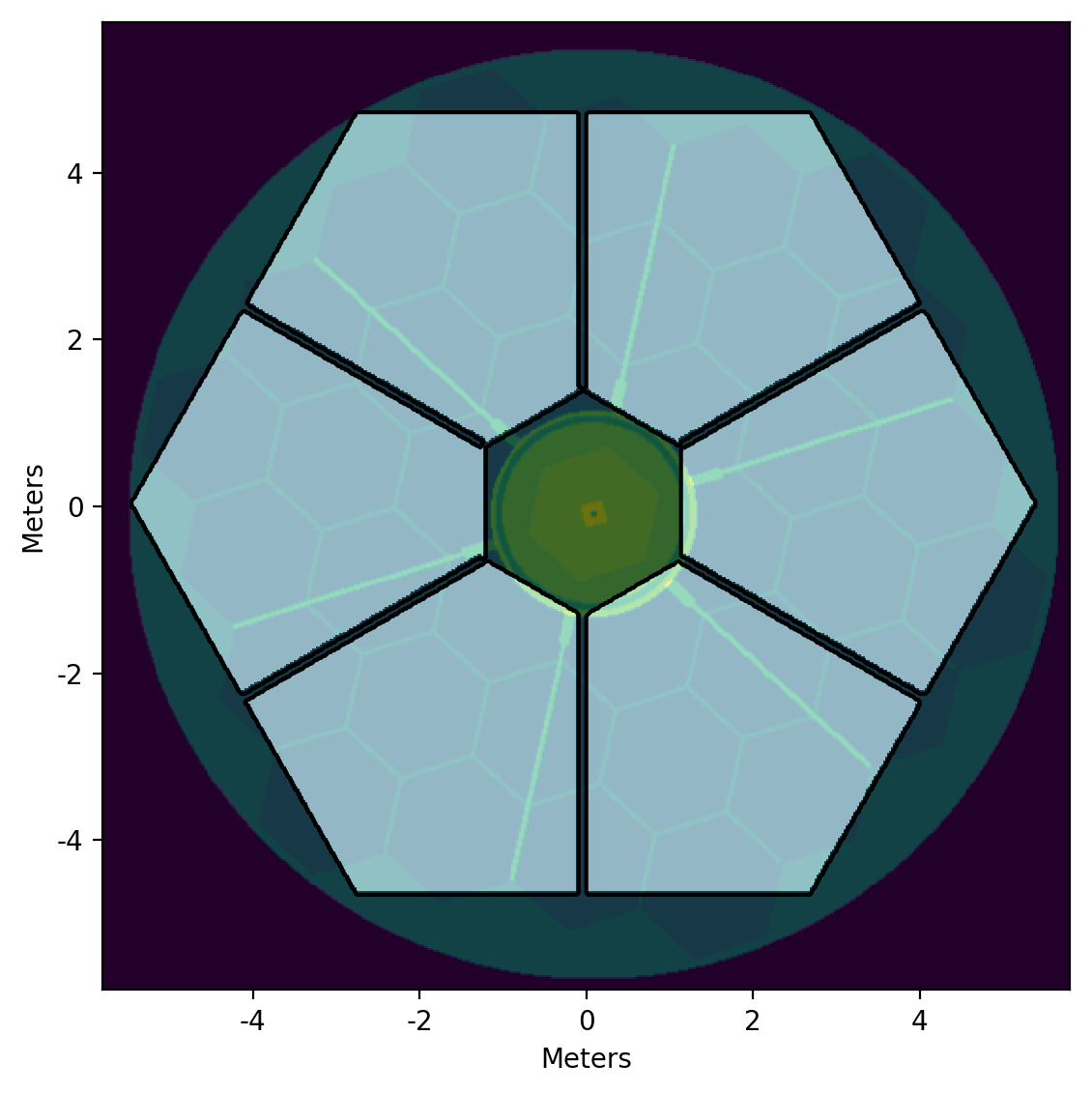}
    \caption{80$^{\circ}$}
\end{subfigure}
\caption{Models of the Keck pupil and background at 4 different elevations with center offsets derived from OSIRIS imaging. The scale of the models are shown in meters projected on the plane of the primary mirror. As one can see, the deviation of pupil centers increases with higher elevation. The rotation of the pupil image after 600s (starting with the mask aligned) is also shown here. The rate of this rotation also increases with elevation.}
\label{fig:pupil_sims}
\end{figure}

The other affect we incorporate into these simulations is nutation of the pupil, which is dependent on the elevation of the telescope and the AO K mirror rotation angle. This results in an offset between the pupil image and the nominal location of the pupil mask. The affect is most pronounced at large elevations where there is a higher rate of field rotation. This diminishes the effectiveness of a rotating pupil mask in the regime where it would otherwise be the most useful. In order to quantify the expected offset between the pupil and mask center, we make use of observations performed with the pupil viewing mode of the OSIRIS imager. Exposures of the telescope pupil without masks in place were taken at a range of telescope elevation and rotator angles. The center of the telescope pupil on the OSIRIS detector is determined from these images and converted to position in cm at the plane of the primary mirror. For elevation angles above 60$^\circ$, we determine an offset of (10, 8.5) cm in (x,y) at the plane of the primary mirror. For elevation angles between 45$^\circ$-60$^\circ$ we apply an offset of (8, 7) cm, and for elevation angles between 30$^\circ$-45$^\circ$ we apply an offset of (6.5, 6) cm. We do not apply any offset below an elevation of 30$^\circ$, though there likely is some. At each elevation, the pupil offset varies with the angle of the AO K mirror, so we use a conservative estimate of the typical offset for inclusion in these simulations.  Figure \ref{fig:pupil_sims} shows the alignment of the ``hex" mask and pupil image at each of these elevation ranges.

\begin{figure}
\begin{subfigure}{0.48\textwidth}
    \centering
    \includegraphics[width=.8\textwidth]{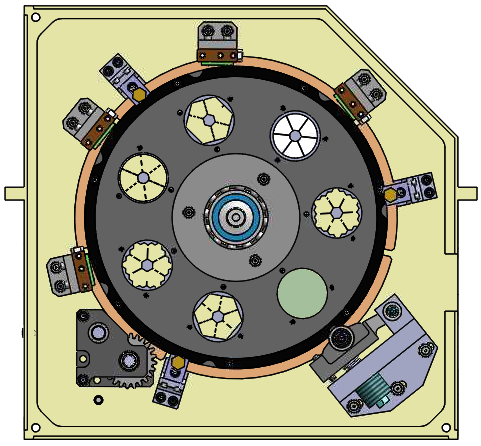}
    \caption{}
\end{subfigure}
\begin{subfigure}{0.48\textwidth}
    \centering
    \includegraphics[width=.8\textwidth]{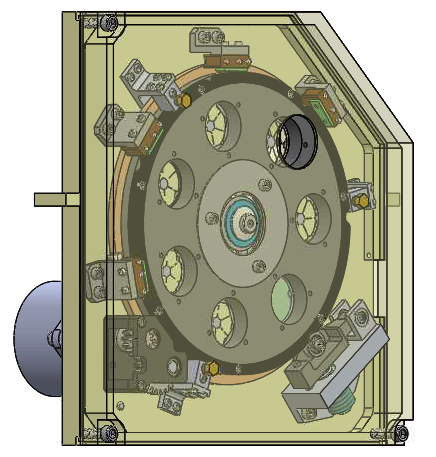}
    \caption{}
\end{subfigure}
\caption{Model views of the current design for the Liger pupil wheel assembly. (a): Front view without the housing to show internal components and sample masks. The same motor as the filter wheels is mounted at the bottom left with a unique spur gear assembly. A single detent is located in the bottom right which is a more compact version of the one in the filter wheel. Four switches and the three spring loaded bearing mechanisms shown in Figure \ref{fig:pupil_bearing} are located around the perimeter of the wheel. Potential standard pupil masks are shown in six of the seven available slots: two each of the annular, hexagonal, and matched hexagon (matched to the edges of the primary mirror) patterns, while a transparent blue aperture denotes a slot for the vector-APP coronagraph. The exact pupil masks that will be used have yet to be determined. (b): Side view of the assembly with the housing included. \label{fig:pupil_wheel}}
\end{figure}

This diminishes the difference in exposure time for the two cases being modelled to what is reported in Table \ref{tbl:backlim_snr} for all elevation angles. Therefore we will proceed with a pupil wheel design in which individual masks \textit{do not} rotate. The current design of the pupil wheel with the fixed masks is shown in Figure \ref{fig:pupil_wheel}.

\section{Summary \& Future Work}
We have presented the latest design of the filter and pupil wheels for the Liger imager and IFS. The full filter wheel will consist of three stacked wheels with 18 filters each (1 clear aperture). These wheels will share a common assembly but will move independently as well as having dedicated switches and detents.

A trade study was performed to determine whether adding a rotating pupil mask would provide significant enough SNR gains to justify the added design complexity. Due to the impacts of pupil nutation and mask misalignment this was determined not to be the case for Liger and we will therefore proceed with the design of a pupil wheel in which individual masks maintain a fixed orientation. 

The pupil wheel is in the early stages of the design, currently containing slots for seven unique pupil masks. The pupil wheel makes use of largely similar components to the filter wheel with a set of binary switches for positioning, the same stepper motor, and a smaller version of the filter wheel detent. Spring loaded bearings are located around the perimeter of the wheel to improve the planarity of the wheel and therefore the repeatability of mask alignment.

Both the filter and pupil wheels will be tested and assembled in a custom test chamber designed for use with Liger (see Wiley et al. (\textit{this conference})\cite{Wiley2020}) which will operate at a temperature below $\rm77 \, K$ and vacuum of $\rm 10^{-5} \, Torr$. 

\acknowledgments
This research program was supported by the Heising-Simons Foundation. We would also like to thank Carlos Alvarez for providing the images of the NIRC2 pupil included in Figure \ref{fig:nirc2_pupil}.

\bibliography{fwheel}

\begin{thebibliography}{1}

\bibitem{Larkin2006}
{Larkin}, J., {Barczys}, M., {Krabbe}, A., {Adkins}, S., {Aliado}, T., {Amico},
  P., {Brims}, G., {Campbell}, R., {Canfield}, J., {Gasaway}, T., {Honey}, A.,
  {Iserlohe}, C., {Johnson}, C., {Kress}, E., {LaFreniere}, D., {Lyke}, J.,
  {Magnone}, K., {Magnone}, N., {McElwain}, M., {Moon}, J., {Quirrenbach}, A.,
  {Skulason}, G., {Song}, I., {Spencer}, M., {Weiss}, J., and {Wright}, S.,
  ``{OSIRIS: a diffraction limited integral field spectrograph for Keck},'' in
  [{\em Society of Photo-Optical Instrumentation Engineers (SPIE) Conference
  Series}{\nolinebreak\hspace{0.1em}]},  {McLean}, I.~S. and {Iye}, M., eds.,
  {\em Society of Photo-Optical Instrumentation Engineers (SPIE) Conference
  Series} {\bf 6269},  62691A (June 2006).

\bibitem{Arriaga2018}
{Arriaga}, P., {Fitzgerald}, M., {Johnson}, C., {Weiss}, J., and {Lyke}, J.~E.,
  ``{Upgrade and characterization of the OSIRIS imager detector},'' in [{\em
  Ground-based and Airborne Instrumentation for Astronomy
  VII}{\nolinebreak\hspace{0.1em}]},  {Evans}, C.~J., {Simard}, L., and
  {Takami}, H., eds., {\em Society of Photo-Optical Instrumentation Engineers
  (SPIE) Conference Series} {\bf 10702},  107022U (July 2018).

\bibitem{Larkin2016}
{Larkin}, J.~E., {Moore}, A.~M., {Wright}, S.~A., {Wincentsen}, J.~E.,
  {Anderson}, D., {Chisholm}, E.~M., {Dekany}, R.~G., {Dunn}, J.~S.,
  {Ellerbroek}, B.~L., {Hayano}, Y., {Phillips}, A.~C., {Simard}, L., {Smith},
  R., {Suzuki}, R., {Weber}, R.~W., {Weiss}, J.~L., and {Zhang}, K., ``{The
  Infrared Imaging Spectrograph (IRIS) for TMT: instrument overview},'' in
  [{\em Ground-based and Airborne Instrumentation for Astronomy
  VI}{\nolinebreak\hspace{0.1em}]},  {Evans}, C.~J., {Simard}, L., and
  {Takami}, H., eds., {\em Society of Photo-Optical Instrumentation Engineers
  (SPIE) Conference Series} {\bf 9908},  99081W (Aug. 2016).

\bibitem{Wright2020}
{Wright}, S.~A., {Larkin}, J.~E., {Jones}, T., {Aliado}, T., {Armus}, L.,
  {Brown}, A., {Chisholm}, E., {Cosens}, M., {Dekaney}, R., {Do}, T.,
  {Fassanacht}, C., {Fisher}, D., {Fitzgerald}, M., {Ghez}, A., {Hirtenstein},
  J., {Johnson}, C., {Kassis}, M., {Keane}, J., {Kelley}, P., {Kirby}, E.,
  {Konopacky}, Q., {Kress}, E., {Kupke}, R., {Lu}, J., {Lyke}, J., {Marley},
  M., {Medling}, A., {Millar-Blanchaer}, M., {Nash}, R., {Nierenberg}, A.,
  {Reddy}, N., {Rich}, M., {Ruffio}, J.-B., {Rundquist}, N.-E., {Sand}, D.,
  {Sanders}, R., {Sandstrom}, K., {Shapley}, A., {Sohn}, J.-M., {Surya}, A.,
  {Treu}, T., {Wang}, E., {Weber}, B., {Wiley}, J., {Wizinowich}, P., {Wong},
  M., {Yeh}, S., and {Zonca}, A., ``{Liger for Next-Generation Keck Adaptive
  Optics: Overall Design and Status},'' in [{\em Ground-based and Airborne
  Instrumentation for Astronomy VI}{\nolinebreak\hspace{0.1em}]},  {Shields},
  J., ed., {\em Society of Photo-Optical Instrumentation Engineers (SPIE)
  Conference Series} {\bf 11447},  11447--331 (Dec. 2020).

\bibitem{shigleys}
Budynas, R.~G. and Nisbett, J.~K.,  [{\em Shigley's Mechanical Engineering
  Design}{\nolinebreak\hspace{0.1em}]}, McGraw-Hill, ninth~ed. (2011).

\bibitem{Snik2012}
{Snik}, F., {Otten}, G., {Kenworthy}, M., {Miskiewicz}, M., {Escuti}, M.,
  {Packham}, C., and {Codona}, J., ``{The vector-APP: a broadband apodizing
  phase plate that yields complementary PSFs},'' in [{\em Modern Technologies
  in Space- and Ground-based Telescopes and Instrumentation
  II}{\nolinebreak\hspace{0.1em}]},  {Navarro}, R., {Cunningham}, C.~R., and
  {Prieto}, E., eds., {\em Society of Photo-Optical Instrumentation Engineers
  (SPIE) Conference Series} {\bf 8450},  84500M (Sept. 2012).

\bibitem{Arriaga16}
{Arriaga}, P., {Fitzgerald}, M.~P., {Lyke}, J.~E., {Campbell}, R.~D.,
  {Wizinowich}, P.~L., {Adkins}, S.~M., and {Matthews}, K.~Y., ``{Modeling the
  transmission and thermal emission in a pupil image behind the Keck II
  adaptive optics system},'' in [{\em Ground-based and Airborne Instrumentation
  for Astronomy VI}{\nolinebreak\hspace{0.1em}]},  {Evans}, C.~J., {Simard},
  L., and {Takami}, H., eds., {\em Society of Photo-Optical Instrumentation
  Engineers (SPIE) Conference Series} {\bf 9908},  990835 (Aug. 2016).

\bibitem{Wiley2020}
{Wiley}, J., {Mathur}, K., {Brown}, A., {Wright}, S., {Cosens}, M., {Maire},
  J., {Fitzgerald}, Michael amd~{Jones}, T., {Kassis}, M., {Kress}, E.,
  {Kupke}, R., {Larkin}, J.~E., {Lyke}, J., {Wang}, E., and {Yey}, S., ``{Liger
  for Next-Generation Keck Adaptive Optics: Cryogenic Chamber for the Imaging
  Camera},'' in [{\em Ground-based and Airborne Instrumentation for Astronomy
  VI}{\nolinebreak\hspace{0.1em}]},  {Shields}, J., ed., {\em Society of
  Photo-Optical Instrumentation Engineers (SPIE) Conference Series} {\bf
  11447},  11447--311 (Dec. 2020).

\end{thebibliography}
\bibliographystyle{spiebib} 

\end{document}